\documentclass[11pt, letterpaper]{article}
\usepackage[margin=1in]{geometry}
\usepackage{booktabs} 
\usepackage[table,xcdraw]{xcolor}
\usepackage{tabularx}
\newcolumntype{Y}{>{\centering\arraybackslash}X} 
\usepackage[sort]{natbib}
\usepackage{url}
\usepackage{graphicx}
\usepackage{color}
\usepackage{setspace}

\usepackage{mathtools}

\usepackage{caption}
\usepackage{longtable}
\newcolumntype{L}{>{\raggedright\arraybackslash}p{2cm}}

\newcommand\numClientsTotal{5,709}
\newcommand\numClientsShort{2,898}
\newcommand\numClientsControl{2,811}
\newcommand\numClientsFirstPhase{2,387}
\newcommand\numClientsSecondPhase{3,322}

\newcommand\firstCourtDateFirstPhase{May 17, 2022}
\newcommand\firstCourtDateSecondPhase{October 14, 2022}
\newcommand\lastCourtDateFirstPhase{September 21, 2022}
\newcommand\lastCourtDateSecondPhase{August 24, 2023}

\newcommand\bwRateControl{12.1\%}
\newcommand\bwRateShort{9.7\%}
\newcommand\bwRateRawPPDiff{2.5pp}
\newcommand\bwRateRawPPCI{0.9pp--4.1pp}
\newcommand\bwRateRawRelativeDiff{20.4\%}

\newcommand\bwAnyRateControl{20.7\%}
\newcommand\bwAnyRateShort{17.6\%}
\newcommand\bwAnyRateRawPPDiff{3.1pp}
\newcommand\bwAnyRateRawPPCI{1.1pp--5.1pp}
\newcommand\bwAnyRateRawRelativeDiff{15.0\%}

\newcommand\incarcerationRateControl{6.2\%}
\newcommand\incarcerationRateShort{4.8\%}
\newcommand\incarcerationRateRawPPDiff{1.4pp}
\newcommand\incarcerationRateRawPPCI{0.2pp--2.5pp}
\newcommand\incarcerationRateRawRelativeDiff{22.0\%}

\newcommand\simpleBWEffectEst{0.775}
\newcommand\simpleBWEffectSE{0.066}
\newcommand\simpleBWEffectStars{**}

\newcommand\simpleAnyBWEffectEst{0.818}
\newcommand\simpleAnyBWEffectSE{0.055}
\newcommand\simpleAnyBWEffectStars{**}

\newcommand\simpleIncarcerationEffectEst{0.769}
\newcommand\simpleIncarcerationEffectSE{0.090}
\newcommand\simpleIncarcerationEffectStars{*}

\newcommand\mainBWEffectEst{0.748}
\newcommand\mainBWEffectSE{0.068}
\newcommand\mainBWEffectStars{**}

\newcommand\mainAnyBWEffectEst{0.797}
\newcommand\mainAnyBWEffectSE{0.058}
\newcommand\mainAnyBWEffectStars{**}

\newcommand\mainIncarcerationEffectEst{0.753}
\newcommand\mainIncarcerationEffectSE{0.095}
\newcommand\mainIncarcerationEffectStars{*}

\newcommand\avgCostPerCase{60¢}

\newcommand\bwRateLongPrereg{13.4\%}
\newcommand\bwRateShortPrereg{12.3\%}

\newcommand\numClientsNoReminderInShort{85}

\newcommand\numOptOut{107}

\newcommand\numOptOutWrongNumber{57}

\newcommand\pctShortConfirmed{51\%}
\newcommand\pctShortUnconfirmed{49\%}
\newcommand\bwRateShortConfirmed{2.9\%}
\newcommand\bwRateShortUnconfirmed{16.8\%}

\newcommand\pctNonEnglish{22\%}

\newcommand\numCrimeTypes{19}

\newcommand\coefficientEstimatesTable{\toprule
\multicolumn{1}{c}{Timeframe} & \multicolumn{1}{c}{First Court Date} & \multicolumn{2}{c}{Any Court Date} \\
\cmidrule(l{3pt}r{3pt}){1-1} \cmidrule(l{3pt}r{3pt}){2-2} \cmidrule(l{3pt}r{3pt}){3-4}
\multicolumn{1}{c}{Outcome} & \multicolumn{2}{c}{Bench Warrant} & \multicolumn{1}{c}{Incarceration} \\
\cmidrule(l{3pt}r{3pt}){1-1} \cmidrule(l{3pt}r{3pt}){2-3} \cmidrule(l{3pt}r{3pt}){4-4}
\multicolumn{1}{c}{Model} & \multicolumn{1}{c}{(1)} & \multicolumn{1}{c}{(3)} & \multicolumn{1}{c}{(5)}\\
\midrule
\addlinespace[0.3em]
\multicolumn{4}{l}{\textbf{Intercept}}\\
\hspace{1em}- & -2.54*** (0.73) & -1.17* (0.58) & -2.91*** (0.87)\\
\addlinespace[0.3em]
\multicolumn{4}{l}{\textbf{Treatment}}\\
\hspace{1em}Reminders & -0.29** (0.09) & -0.23** (0.07) & -0.28* (0.13)\\
\addlinespace[0.3em]
\multicolumn{4}{l}{\textbf{Client race/ethnicity}}\\
\hspace{1em}Asian & -0.26 (0.19) & -0.05 (0.16) & 0.33 (0.27)\\
\hspace{1em}Black & -0.35* (0.17) & -0.07 (0.14) & 0.25 (0.23)\\
\hspace{1em}Hispanic & -0.23 (0.13) & -0.21* (0.11) & 0.13 (0.19)\\
\hspace{1em}Native & 0.57 (0.59) & 0.06 (0.54) & 1.17 (0.79)\\
\hspace{1em}Other & -0.61 (0.36) & -0.12 (0.25) & -0.03 (0.47)\\
\addlinespace[0.3em]
\multicolumn{4}{l}{\textbf{Client information}}\\
\hspace{1em}Is not male & -0.18 (0.12) & -0.07 (0.09) & -0.20 (0.17)\\
\hspace{1em}Age & 0.00 (0.00) & 0.00 (0.00) & 0.00 (0.01)\\
\hspace{1em}Mental health & -0.03 (0.13) & -0.05 (0.10) & 0.26 (0.16)\\
\hspace{1em}New client & -0.07 (0.13) & 0.10 (0.11) & 0.03 (0.21)\\
\hspace{1em}Prefers english & -0.05 (0.13) & 0.06 (0.10) & 0.18 (0.18)\\
\hspace{1em}Years since phone added & 0.54*** (0.11) & 0.36*** (0.09) & -0.10 (0.15)\\
\hspace{1em}Home address recorded & -0.61*** (0.12) & -0.59*** (0.10) & -0.61*** (0.16)\\
\hspace{1em}Miles from home to court & 0.00 (0.00) & 0.00 (0.00) & 0.00 (0.00)\\
\addlinespace[0.3em]
\multicolumn{4}{l}{\textbf{Client history (5 yr counts)}}\\
\hspace{1em}Cases & 0.02* (0.01) & 0.02* (0.01) & 0.02 (0.01)\\
\hspace{1em}Convictions & 0.00 (0.00) & 0.00 (0.00) & 0.00 (0.01)\\
\hspace{1em}1/court dates & -0.05 (0.37) & -0.55 (0.31) & -2.59*** (0.63)\\
\hspace{1em}Bench warrants & 0.01 (0.03) & 0.04 (0.03) & 0.08* (0.03)\\
\hspace{1em}Bench warrants/court dates & 2.87*** (0.45) & 3.07*** (0.39) & 2.89*** (0.63)\\
\addlinespace[0.3em]
\multicolumn{4}{l}{\textbf{Day of week}}\\
\hspace{1em}Monday & 0.64** (0.23) & 0.32 (0.17) & 0.35 (0.28)\\
\hspace{1em}Tuesday & 0.45* (0.22) & 0.27 (0.16) & 0.13 (0.28)\\
\hspace{1em}Wednesday & 0.64** (0.22) & 0.32* (0.16) & 0.40 (0.26)\\
\hspace{1em}Thursday & 0.48* (0.22) & 0.24 (0.16) & 0.12 (0.28)\\
\addlinespace[0.3em]
\multicolumn{4}{l}{\textbf{Month}}\\
\hspace{1em}February & -0.12 (0.24) & 0.00 (0.18) & -0.01 (0.38)\\
\hspace{1em}March & 0.10 (0.25) & 0.04 (0.19) & -0.14 (0.43)\\
\hspace{1em}April & 0.29 (0.26) & -0.12 (0.21) & -0.30 (0.49)\\
\hspace{1em}May & 0.14 (0.23) & -0.16 (0.18) & 0.45 (0.35)\\
\hspace{1em}June & 0.21 (0.21) & -0.09 (0.16) & 0.74* (0.32)\\
\hspace{1em}July & -0.10 (0.22) & -0.50** (0.17) & 0.49 (0.33)\\
\hspace{1em}August & 0.05 (0.21) & -0.90*** (0.17) & 0.34 (0.32)\\
\hspace{1em}September & 0.37 (0.23) & -0.68*** (0.20) & 0.20 (0.37)\\
\hspace{1em}October & 0.40 (0.30) & 0.42 (0.23) & 0.91* (0.40)\\
\hspace{1em}November & 0.07 (0.24) & 0.21 (0.18) & 0.47 (0.37)\\
\hspace{1em}December & 0.06 (0.25) & -0.03 (0.19) & -0.16 (0.41)\\
\addlinespace[0.3em]
\multicolumn{4}{l}{\textbf{Court date info}}\\
\hspace{1em}Appearance number & -0.01 (0.01) & -0.01 (0.01) & -0.04** (0.01)\\
\addlinespace[0.3em]
\multicolumn{4}{l}{\textbf{Case severity}}\\
\hspace{1em}Felony & -0.22 (0.23) & -0.07 (0.18) & -0.06 (0.24)\\
\hspace{1em}Misdemeanor & 0.24 (0.22) & 0.30 (0.18) & -0.04 (0.24)\\
\hspace{1em}Prcs violation & 1.68 (1.27) & 0.05 (0.66) & -0.36 (1.03)\\
\hspace{1em}Probation violation & 1.73 (1.26) & 0.06 (0.61) & 0.65 (0.93)\\
\addlinespace[0.3em]
\multicolumn{4}{l}{\textbf{Courthouse}}\\
\hspace{1em}Hall Of Justice & 0.65 (0.61) & 0.31 (0.49) & 0.03 (0.67)\\
\hspace{1em}Family Court & -1.02 (0.75) & 0.24 (0.53) & 0.57 (0.73)\\
\hspace{1em}Palo Alto & 0.19 (0.62) & -0.25 (0.50) & -0.12 (0.70)\\
\hspace{1em}San Jose Municipal & -0.04 (0.46) & 0.42 (0.34) & -0.15 (0.63)\\
\hspace{1em}Morgan Hill & 0.16 (0.63) & -0.35 (0.51) & 0.06 (0.71)\\
\hspace{1em}Other & 0.40 (0.37) & 0.28 (0.33) & 0.76 (0.46)\\
\addlinespace[0.3em]
\multicolumn{4}{l}{\textbf{Charges}}\\
\hspace{1em}Assault & -0.34** (0.12) & -0.19 (0.10) & 0.43** (0.15)\\
\hspace{1em}Burglary & 0.22 (0.25) & 0.27 (0.20) & 0.44 (0.27)\\
\hspace{1em}Disorderly & -0.15 (0.48) & -0.09 (0.39) & 0.87 (0.48)\\
\hspace{1em}Driving & -0.45*** (0.13) & -0.22* (0.11) & -0.51* (0.20)\\
\hspace{1em}Drugs & 0.54*** (0.13) & 0.56*** (0.12) & 0.47** (0.18)\\
\hspace{1em}Forgery & -0.43 (0.59) & -0.06 (0.47) & -1.72 (1.12)\\
\hspace{1em}Fraud & 0.56** (0.18) & 0.40* (0.16) & 0.01 (0.28)\\
\hspace{1em}Homicide & -1.51 (1.01) & -1.09 (0.60) & -0.61 (1.03)\\
\hspace{1em}Kidnapping & -1.17 (0.60) & -0.87* (0.39) & -0.41 (0.55)\\
\hspace{1em}Larceny & 0.39* (0.16) & 0.53*** (0.13) & 0.36 (0.21)\\
\hspace{1em}Larceny (vehicular) & 0.86*** (0.20) & 0.89*** (0.18) & 0.76** (0.26)\\
\hspace{1em}Probation/parole & -2.20 (1.27) & -0.28 (0.61) & -0.25 (0.93)\\
\hspace{1em}Robbery & 0.01 (0.36) & -0.13 (0.27) & -0.39 (0.44)\\
\hspace{1em}Sex offenses & -1.12* (0.47) & -1.39*** (0.40) & -1.19 (0.74)\\
\hspace{1em}Stolen property & 0.61* (0.24) & 0.75*** (0.22) & 0.12 (0.32)\\
\hspace{1em}Trespassing & 0.58* (0.26) & 0.27 (0.23) & 0.46 (0.33)\\
\hspace{1em}Weapons & -0.17 (0.18) & -0.08 (0.15) & -0.03 (0.24)\\
\hspace{1em}Vandalism & 0.27 (0.18) & 0.30* (0.14) & 0.36 (0.21)\\
\hspace{1em}Other & -0.14 (0.12) & -0.04 (0.10) & 0.35* (0.15)\\
\bottomrule}

\newcommand\sampleStatsTable{\toprule
  & All & Cell on File & Experiment & Control & Treatment\\
\midrule
\addlinespace[0.5em]
\multicolumn{6}{l}{\textbf{Age (years)}}\\
\hspace{1em}18-24 & 12\% & 13\% & 14\% & 13\% & 14\%\\
\hspace{1em}25-34 & 33\% & 34\% & 35\% & 36\% & 34\%\\
\hspace{1em}35-44 & 27\% & 28\% & 27\% & 26\% & 27\%\\
\hspace{1em}45-54 & 15\% & 15\% & 14\% & 14\% & 14\%\\
\hspace{1em}55+ & 13\% & 10\% & 10\% & 11\% & 10\%\\
\addlinespace[0.5em]
\multicolumn{6}{l}{\textbf{Case appearance number}}\\
\hspace{1em}1 & 5\% & 36\% & 3\% & 3\% & 3\%\\
\hspace{1em}2 & 30\% & 16\% & 53\% & 53\% & 52\%\\
\hspace{1em}3-5 & 32\% & 22\% & 33\% & 33\% & 33\%\\
\hspace{1em}6+ & 33\% & 25\% & 11\% & 11\% & 11\%\\
\addlinespace[0.5em]
\multicolumn{6}{l}{\textbf{Case severity}}\\
\hspace{1em}Felony & 35\% & 37\% & 36\% & 36\% & 35\%\\
\hspace{1em}Misdemeanor & 49\% & 57\% & 59\% & 58\% & 59\%\\
\hspace{1em}Supervision Violation & 16\% & 6\% & 6\% & 6\% & 5\%\\
\addlinespace[0.5em]
\multicolumn{6}{l}{\textbf{Courthouse}}\\
\hspace{1em}Hall Of Justice & 67\% & 70\% & 75\% & 75\% & 75\%\\
\hspace{1em}Palo Alto & 12\% & 10\% & 12\% & 12\% & 13\%\\
\hspace{1em}South County & 7\% & 8\% & 9\% & 9\% & 9\%\\
\hspace{1em}Other & 14\% & 11\% & 3\% & 3\% & 3\%\\
\addlinespace[0.5em]
\multicolumn{6}{l}{\textbf{Distance from home to courthouse (miles)}}\\
\hspace{1em}0-0.9 & 6\% & 7\% & 7\% & 7\% & 7\%\\
\hspace{1em}1-3.9 & 19\% & 22\% & 23\% & 22\% & 23\%\\
\hspace{1em}4-7.9 & 24\% & 28\% & 28\% & 29\% & 28\%\\
\hspace{1em}8+ & 26\% & 29\% & 31\% & 31\% & 31\%\\
\hspace{1em}N/A & 25\% & 14\% & 11\% & 11\% & 11\%\\
\addlinespace[0.5em]
\multicolumn{6}{l}{\textbf{Identifies as male}}\\
\hspace{1em}True & 80\% & 80\% & 79\% & 79\% & 78\%\\
\hspace{1em}False & 20\% & 20\% & 21\% & 21\% & 22\%\\
\addlinespace[0.5em]
\multicolumn{6}{l}{\textbf{New client (within prev. year)}}\\
\hspace{1em}True & 34\% & 38\% & 48\% & 48\% & 48\%\\
\hspace{1em}False & 66\% & 62\% & 52\% & 52\% & 52\%\\
\addlinespace[0.5em]
\multicolumn{6}{l}{\textbf{Num. appearances (prev. 5 years)}}\\
\hspace{1em}0 & 4\% & 23\% & 0.5\% & 0.6\% & 0.4\%\\
\hspace{1em}1 & 20\% & 11\% & 37\% & 37\% & 38\%\\
\hspace{1em}2-5 & 26\% & 21\% & 30\% & 30\% & 31\%\\
\hspace{1em}6-19 & 28\% & 24\% & 19\% & 19\% & 18\%\\
\hspace{1em}20+ & 23\% & 22\% & 13\% & 13\% & 13\%\\
\addlinespace[0.5em]
\multicolumn{6}{l}{\textbf{Num. bench warrants (prev. 5 years)}}\\
\hspace{1em}0 & 61\% & 64\% & 70\% & 71\% & 70\%\\
\hspace{1em}1 & 15\% & 13\% & 14\% & 13\% & 14\%\\
\hspace{1em}2-5 & 19\% & 17\% & 13\% & 13\% & 13\%\\
\hspace{1em}6+ & 5\% & 5\% & 3\% & 3\% & 3\%\\
\addlinespace[0.5em]
\multicolumn{6}{l}{\textbf{Potential mental health issue(s)}}\\
\hspace{1em}True & 16\% & 18\% & 15\% & 16\% & 15\%\\
\hspace{1em}False & 84\% & 82\% & 85\% & 84\% & 85\%\\
\addlinespace[0.5em]
\multicolumn{6}{l}{\textbf{Prefers English}}\\
\hspace{1em}True & 82\% & 80\% & 78\% & 78\% & 77\%\\
\hspace{1em}False & 18\% & 20\% & 22\% & 22\% & 23\%\\
\addlinespace[0.5em]
\multicolumn{6}{l}{\textbf{Race and ethnicity}}\\
\hspace{1em}Asian & 8\% & 8\% & 8\% & 8\% & 8\%\\
\hspace{1em}Black & 11\% & 12\% & 11\% & 11\% & 12\%\\
\hspace{1em}Hispanic & 55\% & 60\% & 62\% & 61\% & 62\%\\
\hspace{1em}Native & 0.3\% & 0.4\% & 0.4\% & 0\% & 0.5\%\\
\hspace{1em}White & 17\% & 17\% & 16\% & 16\% & 15\%\\
\hspace{1em}Other & 8\% & 3\% & 3\% & 3\% & 2\%\\
\addlinespace[0.5em]
\textbf{Total clients} & \textbf{21,343} & \textbf{14,063} & \textbf{5,709} & \textbf{2,811} & \textbf{2,898}\\
\bottomrule}

\title{Automated Reminders Reduce Incarceration for Missed Court Dates: Evidence from a Text Message Experiment}

\author{%
\begin{tabular}{cc}
  Alex Chohlas-Wood & Madison Coots \\
  Harvard University & Harvard University \\
  \texttt{achohlaswood@hks.harvard.edu} & \texttt{mcoots@g.harvard.edu} \\[3ex]
  Joe Nudell & Julian Nyarko \\
  Harvard University & Stanford University \\
  \texttt{jnudell@hks.harvard.edu} & \texttt{jnyarko@law.stanford.edu} \\[3ex]
  Emma Brunskill & Todd Rogers \\
  Stanford University & Harvard University \\
  \texttt{ebrun@cs.stanford.edu} & \texttt{todd\_rogers@hks.harvard.edu} \\[3ex]
  \multicolumn{2}{c}{Sharad Goel} \\
  \multicolumn{2}{c}{Harvard University} \\
  \multicolumn{2}{c}{\texttt{sgoel@hks.harvard.edu}} \\
\end{tabular}
}

\date{}

\begin{document}
\pagenumbering{gobble}
\maketitle

\begin{abstract}
\noindent 
Millions of Americans 
must attend mandatory court dates every year.
To boost appearance rates,
jurisdictions nationwide are increasingly turning to automated reminders,
but previous research offers mixed evidence 
on their effectiveness. 
In partnership with the Santa Clara County Public Defender Office,
we randomly assigned \numClientsTotal{} public defender clients to either 
receive automated text message reminders (treatment)
or not receive 
reminders (control).
We found that reminders reduced warrants issued for missed court dates by approximately
20\%,
with \bwRateControl{} of clients in the control condition issued a warrant
compared to \bwRateShort{} of clients in the treatment condition.
We further found that incarceration from missed court dates dropped by a similar amount, 
from \incarcerationRateControl{} in the control condition to \incarcerationRateShort{} in the treatment condition.
Our results provide evidence that automated reminders can help people avoid the negative consequences of missing court.
\end{abstract}

\clearpage

\pagenumbering{arabic}
\setcounter{page}{1}
\onehalfspacing
\section{Introduction}

In the United States,
after a person is arrested and charged with a crime,
they are either held in jail as their case proceeds,
or they are released and asked to attend court of their own accord.
While many released defendants do indeed attend court---%
as is legally required---%
some fail to do so.
Non-appearance rates vary depending on jurisdiction and offense type, 
ranging from less than 10\% to as high as 50\%~\citep{bornstein2013reducing,owens2023can}.
Failing to appear (FTA) at a required court date is a crime in nearly every state,
and non-appearance can prompt judges to issue a warrant
mandating the defendant's arrest---hereafter called a ``bench warrant''---at their next 
encounter with law enforcement~\citep{ncsl2018}.
Missed court dates can also create inefficiencies for the court system more broadly, which
increasing costs and exacerbating post-pandemic delays in U.S. courts~\citep{jurva2021impacts}.

Once arrested for a bench warrant, 
punishment can include time in jail.
This pretrial incarceration 
can impose social and economic hardships on defendants and their families---including housing loss, family strain, and social stigma~\citep{bergin2022initial}---even as evidence suggests that time in jail may not deter future court absences~\citep{holsinger2023pretrial}.
Pretrial incarceration has also been found to increase recidivism and reduce employment~\citep{dobbie2018effects,loeffler2022impact,smith2022pretrial}.
The consequences of missed court dates may fall particularly hard on racial minorities, 
given the disproportionate involvement of marginalized communities in the criminal legal system~\citep{pew2022disparities}.

People may miss their court date simply due to forgetfulness or confusion about the court system~\citep{kofman_2019}. 
As a result, court date reminders  
are increasingly used
to help people remember and plan for
their upcoming court obligations.
Nearly half of all counties nationwide
have either implemented or are planning to implement
court date reminders via text message, phone call, mail, or some other method~\citep{lattimore2020prevalence}.
Yet research on the effects of automated text message reminders is limited, 
even though it is one of the most cost-effective ways of sending reminders,
and is increasingly used by jurisdictions nationwide.\footnote{There is a larger literature on the effectiveness of
court date reminders by mail
or telephone call~\citep{crozier2000court,goldkamp2006restoring,howat2016improving,schnacke2012increasing,ferri2020benefits,nice2006court,foudray2022jail,white2006court,tomkins2012experiment},
and on the effectiveness of text message reminders to other participants in the criminal legal system
\citep{cumberbatch2018nudge,hastings2021reducing}.
For example, in an experiment in Arkansas, \citet{hastings2021reducing} found that text message reminders reduced missed probation and parole appointments by over 40\%,
and \citet{tomkins2012experiment} found that postcard reminders reduced non-appearance rates by up to 34\% 
in an experiment with
misdemeanor defendants in Nebraska.
See \citet{bechtel2017meta} and \citet{zottola2023court} for reviews of the relevant literature.
}
The literature that does exist paints an incomplete picture
of the efficacy of text message reminders to
increase court appearance and decrease the negative consequences of missing court (Table~\ref{tab:lit_review}).
Two recent randomized controlled trials (RCTs) found significant and meaningful reductions in FTA rates from text message reminders~\citep{emanuel2022tripping,fishbane2020behavioral},
while two other RCTs also found reductions in FTA rates, 
though the estimates were not statistically significant~\citep{lowenkamp2018assessing,owens2023can}.
\cite{emanuel2022tripping} additionally examined the impact of automated reminders on pretrial incarceration, finding no statistically significant effect of reminders on jail bookings. One RCT estimated \emph{higher}---but not statistically significant---warrant rates 
among people who received a text message reminder~\citep{chivers2018sorry}.
Some of these studies likely lacked sufficient statistical power,
but their presence in the literature may confuse or deter policymakers who are already hesitant to adopt behavioral nudges~\citep{dellavigna2023bottlenecks}.
In addition, we note that the two studies that found statistically significant impacts on FTA 
considered only municipal violations and misdemeanors, even though more serious felony-level cases often comprise a substantial proportion of a court's caseload.

\begin{table}[t!]
\renewcommand{\arraystretch}{2.5}
\tiny
\hyphenpenalty10000
\begin{tabularx}{\textwidth}{>{\raggedright}Xl>{\raggedright}X>{\raggedright}XlllXXXXX}
\toprule
\textbf{Study}              & \textbf{Year} & \textbf{Outcome}              & \textbf{Sample}      & \textbf{Control} & \textbf{Est. effect} & \textbf{CI}  & \textbf{Estimated rel.~effect} & \textbf{P-val}  \\ \midrule
Chivers \& Barnes, 2018     & 2017          & Warrant at court date           & 946 defendants            & 22.5\%                  & +1.8pp                 & N/A                 & + 8\% & 0.51                                        \\
Lowenkamp et al., 2018      & N/A            & FTA at court date              & 10,228 defendants         & 13\%               & -2pp               & N/A                & -18\% & 0.07                                      \\
Fishbane et al., 2019       & 2016--17    & FTA/warrant at summons hearing         & 20,234 defendants         & 37.9\%             & -9.9pp             & {[}-12~--~-7.8pp{]} & -26\% & $<$0.01                 \\
Emanuel \& Ho, 2022         & 2018--19    & FTA at arraignment  & 30,870 defendants         & 21\%             & -8.2pp             & N/A  & -39\% & $<$0.01                                  \\
Owens \& Sloan, 2023         & 2021           & FTA at court date              & 1,096 housed defendants   & 50\%                & -6pp               & {[}-11.2~--~+0.6pp{]}               & -12\% & 0.08  \\ 
\bottomrule
\end{tabularx}
\caption{Past experiments have yielded mixed results on the effectiveness of text message court date reminders for improving appearance rates.}
\label{tab:lit_review}
\end{table}

We aim to help resolve whether text message reminders increase court appearance and reduce incarceration.
To do so, we ran an RCT with \numClientsTotal{} clients 
of the Santa Clara County Public Defender Office (SCCPDO),
headquartered in San Jose, California, who were charged with felonies, misdemeanors, or supervision violations.\footnote{
Our pre-registration is available at \url{https://aspredicted.org/SMY_N1R}.
In our original design, we proposed comparing two message variants: 
a variant drafted by the public defender versus
a simpler variant that we thought would be easier for clients to understand.
However, we later concluded that the two message variants were not meaningfully comparable and so terminated that experiment without analyzing any of the resulting data.
Simultaneously, we ran the experiment described in this paper, comparing the simpler variant against no messages. 
For transparency, we have now analyzed the data corresponding to our two-variant experiment, 
finding that bench warrant rates were lower among clients receiving the simpler variant compared to the longer one (\bwRateShortPrereg{} vs. \bwRateLongPrereg{}, respectively), 
although the difference was not statistically significant.
We are currently running a new experiment that we believe is better designed to compare differing message templates, pre-registered at \url{https://aspredicted.org/FKC_XYY}.
}
Alongside our contribution to the general literature on text message court date reminders, 
our study is the first to specifically examine the effect of reminders for clients of a public defender.
Understanding the efficacy of reminders for this subpopulation is particularly important for ongoing policy debates, 
as policymakers may expect that public defenders can ensure court appearance for their own clients, 
obviating the need for reminders sent at additional cost to taxpayers.
Indeed, SCCPDO clients appear at their court appointments the vast majority of the time.
Yet there is still room for improvement,
with about 10--15\% of scheduled court dates for SCCPDO clients ending in a bench warrant for non-appearance. 
Given that people are often required to attend multiple court dates, 
nearly one-third of SCCPDO clients received at least one bench warrant for missing court over the course of 2022.
Over half of these clients 
were only facing misdemeanor or lesser charges,
and one out of every four 
had no history of prior charges
on file with SCCPDO.
A single bench warrant for these clients
thus has the potential to quickly ramp up 
an otherwise minimal brush
with the criminal legal system,
and underscores the importance of increasing appearance rates.

\section{Experiment Design}

Our experiment consists of \numClientsTotal{} SCCPDO clients 
who had court dates during two timespans:
\numClientsFirstPhase{} clients
between \firstCourtDateFirstPhase{} and \lastCourtDateFirstPhase{}, 
and \numClientsSecondPhase{} clients between \firstCourtDateSecondPhase{} and \lastCourtDateSecondPhase{}. 
To be eligible for inclusion in the experiment,
clients 
must have had at least one court date in the timespans mentioned above,
had a cellphone number available in SCCPDO's case management system,
and had never previously received an automated reminder from SCCPDO.\footnote{
We briefly paused our experiment in between the two time periods while we updated our text message delivery system, as discussed in the Appendix. 
Starting in 2021, 
as we developed the software necessary to conduct this experiment, 
we sent court date reminders to some SCCPDO clients; 
these clients were not eligible for inclusion in our experiment,
though they received similar reminders to those described here.
}

We focus on two primary outcome metrics:
(1)
issuance of a bench warrant for failure-to-appear (FTA)
after assignment to treatment or control;
and (2)
whether a client was remanded to custody
on a bench warrant 
at any point between assignment and the end of the experiment.
Judges often issue a bench warrant when a defendant does not attend a mandatory court date, 
though they can decline to do so if they believe the client has sufficient justification for not being present 
(e.g., being sick with COVID)~\citep{graef2023consequences}. 
We consider both whether a bench warrant was issued at a client's first scheduled court date,
as well as whether clients received at least one bench warrant 
at any time during the experiment.

Bench warrants are rarely proactively enforced.
In theory, clients may resolve an open bench warrant
by independently reaching out to the court or their attorney.
In practice, however,
clients with open bench warrants are typically arrested 
during their next, unrelated encounter with law enforcement (e.g., as a result of a traffic stop).
In either case,
the client then appears at a ``bench warrant hearing.''
At these hearings,
judges may choose to release the client back into the community if they believe the client will appear at future court dates,
or---alternatively---may 
remand the client to jail pending bail, later release, or case resolution.
As result, 
a bench warrant does not always result in incarceration,
given the dynamics of client initiative,
chance contact with law enforcement,
and judicial discretion.
In particular,
only half of SCCPDO clients with bench warrants issued between 2019 and 2021 
were remanded to jail 
within two years of their missed court date (Figure~\ref{fig:time-to-remand}).

We are specifically interested in how court date reminders impact these remands to jail.
To measure this phenomenon, 
we code a client's outcome as ``incarcerated'' if they were remanded at a bench warrant hearing where no new charges were brought,
and code the outcome as ``not incarcerated'' for all other clients.~\footnote{
To verify that a client who was remanded to custody was, in fact, 
held in the county jail,
we worked with SCCPDO to manually query the custody status 
for a sample of 41 clients with court dates between June 5 and June 7, 2023.
Of these 41 clients, 18 were remanded to custody at their court date. 
As of June 9, 2023, 16 of these 18 clients remanded to custody were verified to be in jail.
Of the remaining 23 clients (who were not remanded to custody), 22 were verified not to be in jail.
The small discrepancy between remands and incarceration is likely due to events that transpired between the court date and the custody check; 
for example, clients may have been released after paying bail,
or may have been incarcerated on a different case not represented by SCCPDO.
These results suggest that the vast majority of clients remanded to custody spend at least several days in jail.
}
This metric directly corresponds to the target of our intervention---incarceration solely attributable to missed court dates.
Our findings are qualitatively similar if we redefine the outcome to indicate whether a 
client was remanded at any type of bench warrant hearing, 
regardless of whether they were arrested on new charges.

\begin{figure}[t]
\begin{center}
\centerline{\includegraphics[width=\columnwidth]{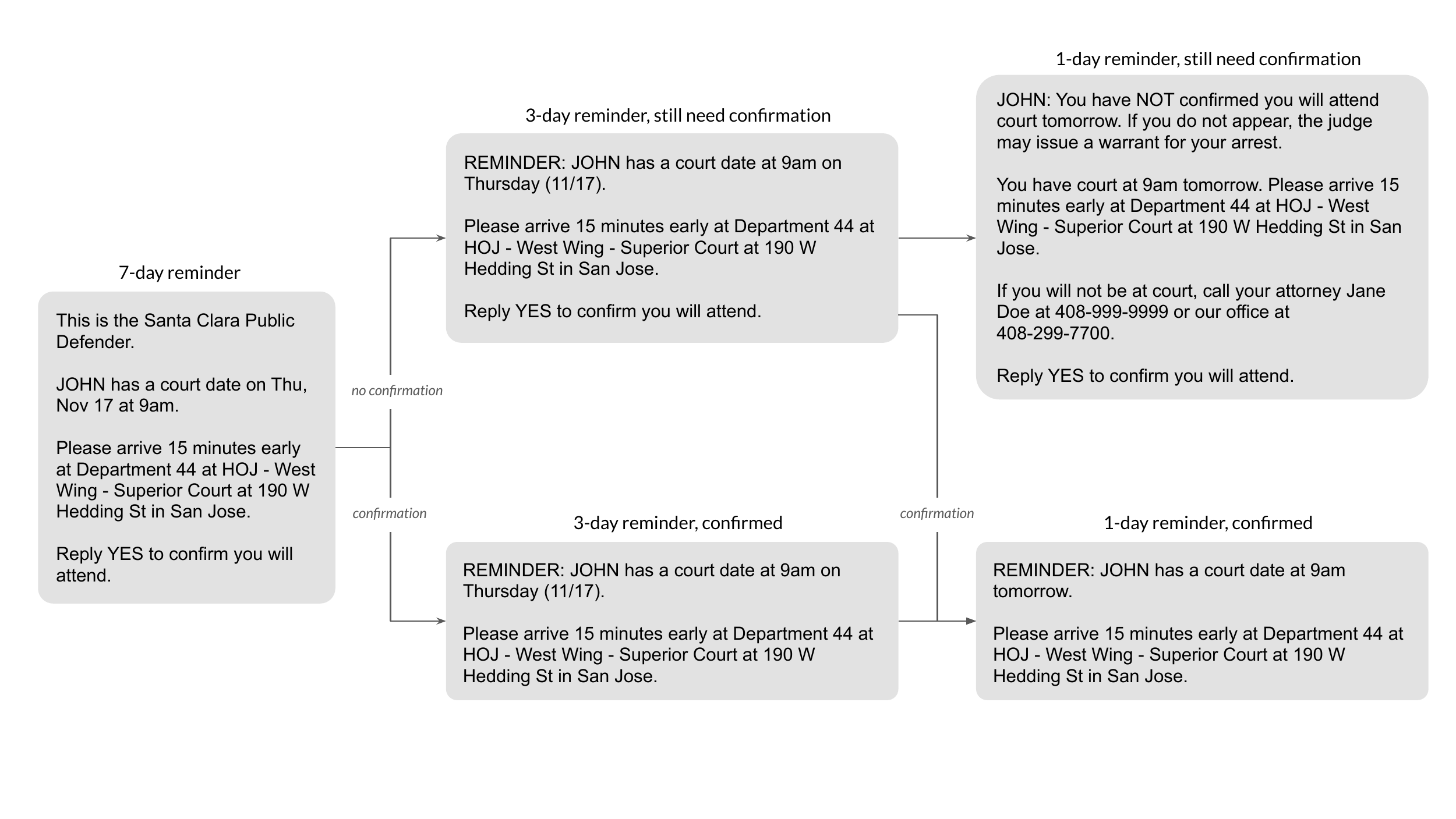}}
\vspace{-2em}
\caption{
Message flow for clients in the treatment condition.
Clients are asked to confirm their attendance at each court date,
with the timing of their confirmation determining their path through this flow.
For example, a client who confirms immediately after the first reminder would follow the bottom path.
Other clients who withhold any confirmation
would follow the top path.
}
\vspace{-2em}
\label{fig:treatment-diagram-english}
\end{center}
\end{figure}

\begin{figure}[t]
    \centering
    \includegraphics[width=0.8\textwidth]{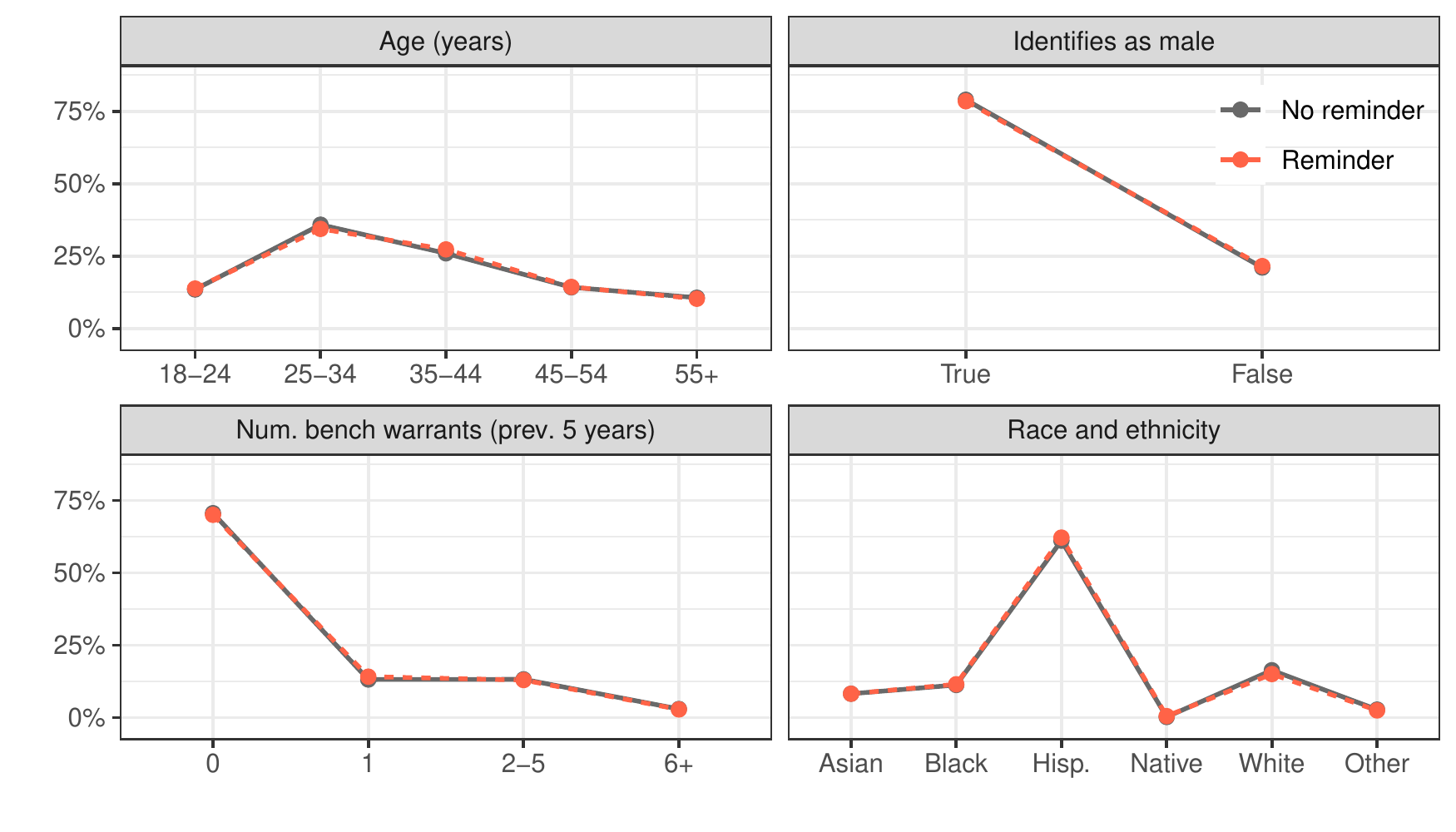}
    \vspace{-0.8em}
    \caption{Distributions of select covariates split between treatment and control populations. 
    Nearly half of participants were under the age of 34,
    most identified as male, 
    a majority identified as Hispanic,
    and most had no record of a bench warrant within the past five years.
    In addition, 
    distributions are nearly identical between conditions, confirming that our assignment mechanism randomly assigned clients to the two conditions as intended. 
    See Figure~\ref{fig:full_balance_plot} for a more comprehensive version of this plot.
    }
    \label{fig:limited_balance_plot}
\end{figure}

The \numClientsTotal{} SCCPDO clients in our experiment were randomly assigned to treatment or control conditions with equal probability:
\numClientsControl{} clients were assigned to the control condition, 
which meant they did not receive any automated reminders;
and \numClientsShort{} clients were assigned to the treatment condition,
which meant they received a series of automated reminders before their court date.
In Figures~\ref{fig:limited_balance_plot} and \ref{fig:full_balance_plot}, 
we show that covariate distributions were nearly identical
across experiment arms,
indicating that the randomization scheme worked as intended.

Prior to the first reminder, we sent an introductory text message to clients in the treatment condition explaining the reminder program
and explaining how to opt out, if desired.
Of the \numClientsShort{} clients in the treatment arm, \numOptOut{} opted out of receiving text message reminders,
of which a majority (\numOptOutWrongNumber{} clients) opted out by noting that we had the wrong number.
Reminders began seven days before each upcoming court date,
with another reminder three days before, 
and a final reminder the day before the court date.
(See Figure~\ref{fig:treatment-diagram-english} for a diagram of these reminders.)
This schedule mirrors the timing of reminders in \citet{fishbane2020behavioral}, and is similar to the timing used in other studies, which sent reminders at various combinations of one, three, and seven days before a court date~\citep{chivers2018sorry,emanuel2022tripping,ferri2020benefits,zottola2023court}. 
Translated versions of these reminders were provided in Spanish and Vietnamese for the \pctNonEnglish{} of clients who had previously indicated a need for a translator in one of these languages (Figures~\ref{fig:treatment-diagram-spanish} and~\ref{fig:treatment-diagram-vietnamese}).

Clients were prompted to confirm their attendance by responding with ``yes'' or similar affirmations.
For example, our application recognized many possible confirmations, including ``OK'', ``Confirmed'', ``I'll be there'', and a thumbs-up emoji, and confirmations in Spanish and Vietnamese.
If they confirmed,
we did not prompt for confirmation on subsequent reminders.
Ultimately, \pctShortConfirmed{} of clients in the treatment arm confirmed their attendance,
and among these clients, 
\bwRateShortConfirmed{} received a bench warrant at their first court date;
in comparison, a bench warrant was issued for \bwRateShortUnconfirmed{} of clients who did not confirm their attendance (Table~\ref{tab:confirmation_and_appearance_behavior}).
This difference could be explained by the act of confirming,
self-selection,
or a combination thereof.
In any case,
our study is not designed to determine whether confirmations affect appearance in court,
and we do not consider confirmation behavior in our analysis
in order to avoid post-treatment bias.

\section{Results}
\label{sec:results}

\begin{table}[t]
{
\renewcommand{\arraystretch}{1.2}
\centering
\begin{tabularx}{\linewidth}{l *{6}{Y}}
\toprule\midrule[0.4pt]
Timeframe & \multicolumn{2}{c}{First court date} & \multicolumn{4}{c}{Any court date} \\ 
 \cmidrule(l){2-3} \cmidrule(l){4-7}
Outcome & \multicolumn{4}{c}{Bench warrant} & \multicolumn{2}{c}{Remand to custody} \\ 
 \cmidrule(l){2-5} \cmidrule(l){6-7}

& (1) & (2) & (3) & (4) & (5) & (6) \\
\midrule
Observations & \multicolumn{2}{c}{ \numClientsTotal{} clients} & \multicolumn{2}{c}{ \numClientsTotal{} clients} & \multicolumn{2}{c}{ \numClientsTotal{} clients} \\
Obs. (control) & \multicolumn{2}{c}{ \numClientsControl{} clients} & \multicolumn{2}{c}{ \numClientsControl{} clients} & \multicolumn{2}{c}{ \numClientsControl{} clients} \\
Obs. (treatment) & \multicolumn{2}{c}{ \numClientsShort{} clients} & \multicolumn{2}{c}{ \numClientsShort{} clients} & \multicolumn{2}{c}{ \numClientsShort{} clients} \\
Rate (control) & \multicolumn{2}{c}{ \bwRateControl{} } & \multicolumn{2}{c}{ \bwAnyRateControl{} } & \multicolumn{2}{c}{ \incarcerationRateControl{} } \\
Rate (treatment) & \multicolumn{2}{c}{ \bwRateShort{} } & \multicolumn{2}{c}{ \bwAnyRateShort{} } & \multicolumn{2}{c}{ \incarcerationRateShort{} } \\
Difference & \multicolumn{2}{c}{ -\bwRateRawPPDiff{} } & \multicolumn{2}{c}{ -\bwAnyRateRawPPDiff{} } & \multicolumn{2}{c}{ -\incarcerationRateRawPPDiff{} } \\
Est. treat. effect & 
\mainBWEffectEst{}\mainBWEffectStars{}  &  
\simpleBWEffectEst{}\simpleBWEffectStars{} & \mainAnyBWEffectEst{}\mainAnyBWEffectStars{}  &  \simpleAnyBWEffectEst{}\simpleAnyBWEffectStars{} &  \mainIncarcerationEffectEst{}\mainIncarcerationEffectStars{} &  \simpleIncarcerationEffectEst{}\simpleIncarcerationEffectStars{} \\
(Std. Error) & (\mainBWEffectSE{}) & (\simpleBWEffectSE{}) & (\mainAnyBWEffectSE{}) & (\simpleAnyBWEffectSE{}) & (\mainIncarcerationEffectSE{}) & (\simpleIncarcerationEffectSE{}) \\
Covar. adjustment & Yes & \multicolumn{1}{c}{No} & Yes & \multicolumn{1}{c}{No} & Yes & \multicolumn{1}{c}{No} \\
\bottomrule
\end{tabularx}
\caption{The effect of text message reminders on the issuance of bench warrants for non-appearance, and on remanding to custody on a bench warrant, estimated using logistic regression as discussed in Section~\ref{sec:results}. 
Reported estimates are odds ratios (i.e., exponentiated logistic regression coefficients),
with standard errors in parentheses calculated using the delta method.
The single star indicates that the corresponding logistic regression coefficient estimates (on the log-odds scale) have a p-value
between 0.01 and 0.05,
and the double star indicates a p-value between 0.001 and 0.01.
Coefficient estimates for other covariates are included in Table~\ref{tab:model_coefs}.
}
\label{tab:regression_results}
}
\end{table}

In the control condition, 
\bwRateControl{} of clients
received a bench warrant
at their first scheduled court date during our experiment window,
compared to \bwRateShort{} for clients in the treatment condition.
This \bwRateRawPPDiff{} difference (95\% CI: \bwRateRawPPCI{}) 
corresponds to a \bwRateRawRelativeDiff{} reduction in bench warrant rates. 
Over the course of the entire experiment, \bwAnyRateControl{} of clients in the control condition received at least one bench warrant,
compared to \bwAnyRateShort{} of clients in the treatment condition,
a \bwAnyRateRawPPDiff{} difference (95\% CI: \bwAnyRateRawPPCI{}) corresponding to a \bwAnyRateRawRelativeDiff{} reduction in the issuance of bench warrants.
These reductions persist in subsequent rates of incarceration.
\incarcerationRateControl{} of clients in the control condition were remanded on a bench warrant after assignment to our experiment, 
compared to \incarcerationRateShort{} of clients in the treatment condition,
a \incarcerationRateRawPPDiff{} difference (95\% CI: \incarcerationRateRawPPCI{})
corresponding to a relative reduction of \incarcerationRateRawRelativeDiff{}.

To improve the precision of our results, we also estimate the impact of text message reminders
via logistic regression models corresponding to each of our outcomes of interest:
\begin{equation}
\Pr(Y_i=1) = \text{logit}^{-1}(\alpha + \beta T_i + \gamma^T X_i),
\end{equation}
where $Y_i$ indicates our outcome of interest (e.g., issuance of a bench warrant),
$T_i$ indicates whether the client was in the treatment condition,
and $X_i$ is a vector representing
a variety of observable features of the client, case,
and first scheduled court date.
In particular, $X_i$ includes:
demographic information 
(%
the client's 
age, 
race, 
whether the client identifies as male, 
whether the client prefers a language interpreter, 
whether the client's attorney indicated a possible mental health issue for the client, 
whether a home address is on file for the client,
and the distance between the client’s home address and the courthouse where their appearance is scheduled, coded as zero if there is no address on file%
);
client history 
(%
the number of bench warrants for non-appearance known to SCCPDO in the previous five years, 
the inverse number of court dates known to SCCPDO in the previous five years,
the product of these two covariates, representing the client's bench warrant rate for failing to appear over the last five years,
whether the client was ``new'', i.e., whether the earliest court date known to the public defender was in the preceding year,
the number of previous cases with the public defender's office,
the number of previous convictions or guilty pleas with the public defender office (including \textit{nolo contendere} pleas),
and 
the number of years since the client's phone records were updated%
);
case information 
(whether the most serious charge was classified as a felony, misdemeanor, or supervision violation,
 and indicators for which of \numCrimeTypes{} high-level charge categories were present, e.g., burglary or robbery);
and court date information 
(%
the courthouse where the court date was scheduled, 
the day of week, 
the month,
and a number indicating the court date was the $n$-th scheduled appointment on a case%
).
With these models, 
we estimate effects that are comparable to those seen with the raw, unadjusted rates (Table~\ref{tab:regression_results}).

\section{Conclusion}
Prisons and jails in the United States are overcrowded and underresourced~\citep{sacbee}, 
and arrests stemming from missed court dates are a significant contributor to incarceration. 
As states attempt to reduce the number of people they incarcerate\footnote{For example, the Supreme Court of the United States ordered California to reduce the size of its prison population because overcrowding rendered prison conditions unconstitutional (see \textit{Brown v. Plata}, 2011, no. 09-12330).}, 
many are looking to court reminders as a way to increase court appearances and reduce jail time.
With an average marginal cost of roughly \avgCostPerCase{} per defendant per case, our results suggest that a text message reminder program can be an effective and relatively inexpensive way to increase appearances and decrease incarceration.

Much remains unanswered about how to design behavioral nudges to be most effective at preventing bench warrants.
For example, the optimal timing and frequency of text message reminders is unclear.
It may be more effective to remind clients about court obligations over a week in advance
or to do so more frequently in the week before.
The reminders we used also only briefly mentioned the possible consequences of missing court.
Other content---a stronger focus on the consequences, or a focus on possible supports---may be more effective at preventing bench warrants.
In addition, court date reminders may not help clients who are struggling with more fundamental barriers to court attendance, such as lack of transportation or childcare, or work obligations.
Other behavioral nudges like transportation or financial assistance might further address these barriers and could complement court date reminders~\citep{brough2022can}.
Other court participants (including witnesses and police officers) 
may also struggle to attend court,
and may benefit from reminders like the ones we describe here~\citep{graef2023systemic}. 
Finally, many public defender clients may lack a reliable and persistent cellphone number altogether, preventing them from benefiting from these types of interventions.

In addition to behavioral nudges, 
policymakers 
might consider alternate pathways to reducing pretrial incarceration.
For example, judges could issue a bench warrant for non-appearance only in the most egregious circumstances, 
such as when there is clear evidence a defendant is unwilling to cooperate with the judicial process.
Some counties in California are working to improve appearance rates and other outcomes by 
pairing defendants with case managers
that help to address underlying challenges, like housing instability and substance use, that their clients may be facing.
While our work demonstrates the promise of behavioral nudges for reducing incarceration, this approach is but one step
in more broadly reforming the criminal legal system.

\section*{Acknowledgements}
We thank our partners at Santa Clara County, including Molly O'Neal, Sarah McCarthy, Terrence Charles, Sven Bouapha, Charlie Hendrickson, Srini Musunuri, and Angel Chan 
for their efforts on this project.
Sophie Allen informed numerous aspects of this study through her fieldwork in Santa Clara County,
and we are  grateful for her continued perspective.
Many other colleagues made valuable contributions to this work,
including:
Ro Encarnacion, Amelia Goodman, Dan Jenson, Nancy Mandujano, Ayesha Omarali;
as well as Tara Watford, Chris Correa, and others from The Bail Project.
This research was supported by grants from Stanford Impact Labs, Stanford Law School, Stanford Community Engagement, and the Harvard Data Science Initiative.

\singlespacing
\bibliography{refs}

\begin{thebibliography}{}

\bibitem[Bechtel et~al., 2017]{bechtel2017meta}
Bechtel, K., Holsinger, A.~M., Lowenkamp, C.~T., and Warren, M.~J. (2017).
\newblock A meta-analytic review of pretrial research: {R}isk assessment, bond
  type, and interventions.
\newblock {\em American Journal of Criminal Justice}, 42:443--467.

\bibitem[Bergin et~al., 2022]{bergin2022initial}
Bergin, T., Ropac, R., Randolph, I., and Joseph, H. (2022).
\newblock The initial collateral consequences of pretrial detention:
  Employment, residential stability, and family relationships.
\newblock {\em Residential Stability, and Family Relationships (September 12,
  2022)}.

\bibitem[Bornstein et~al., 2013]{bornstein2013reducing}
Bornstein, B.~H., Tomkins, A.~J., Neeley, E.~M., Herian, M.~N., and Hamm, J.~A.
  (2013).
\newblock Reducing courts' failure-to-appear rate by written reminders.
\newblock {\em Psychology, Public Policy, and Law}, 19(1):70.

\bibitem[Brough et~al., 2022]{brough2022can}
Brough, R., Freedman, M., Ho, D.~E., and Phillips, D.~C. (2022).
\newblock Can transportation subsidies reduce failures to appear in criminal
  court? {E}vidence from a pilot randomized controlled trial.
\newblock {\em Economics Letters}, 216:110540.

\bibitem[Chivers and Barnes, 2018]{chivers2018sorry}
Chivers, B. and Barnes, G. (2018).
\newblock Sorry, wrong number: Tracking court attendance targeting through
  testing a “nudge” text.
\newblock {\em Cambridge Journal of Evidence-Based Policing}, 2:4--34.

\bibitem[Crozier, 2000]{crozier2000court}
Crozier, T.~L. (2000).
\newblock The court hearing reminder project: If you call them, they will come.
\newblock Technical report, King County District Court.

\bibitem[Cumberbatch and Barnes, 2018]{cumberbatch2018nudge}
Cumberbatch, J.~R. and Barnes, G.~C. (2018).
\newblock This nudge was not enough: A randomised trial of text message
  reminders of court dates to victims and witnesses.
\newblock {\em Cambridge Journal of Evidence-Based Policing}, 2:35--51.

\bibitem[DellaVigna et~al., 2023]{dellavigna2023bottlenecks}
DellaVigna, S., Kim, W., and Linos, E. (2023).
\newblock Bottlenecks for evidence adoption.
\newblock {\em Journal of Political Economy}.
\newblock (Forthcoming).

\bibitem[Dobbie et~al., 2018]{dobbie2018effects}
Dobbie, W., Goldin, J., and Yang, C.~S. (2018).
\newblock The effects of pre-trial detention on conviction, future crime, and
  employment: Evidence from randomly assigned judges.
\newblock {\em American Economic Review}, 108(2):201--240.

\bibitem[Emanuel and Ho, 2022]{emanuel2022tripping}
Emanuel, N. and Ho, H. (2022).
\newblock Tripping through hoops: The effect of violating compulsory government
  procedures.
\newblock {\em American Economic Journal: Economic Policy}.
\newblock Forthcoming.

\bibitem[Ferri, 2022]{ferri2020benefits}
Ferri, R. (2022).
\newblock The benefits of live court date reminder phone calls during pretrial
  case processing.
\newblock {\em Journal of Experimental Criminology}, 18:149–--169.

\bibitem[Fishbane et~al., 2020]{fishbane2020behavioral}
Fishbane, A., Ouss, A., and Shah, A.~K. (2020).
\newblock Behavioral nudges reduce failure to appear for court.
\newblock {\em Science}, 370(6517):eabb6591.

\bibitem[Foudray et~al., 2022]{foudray2022jail}
Foudray, C.~M., Lawson, S.~G., and Lowder, E.~M. (2022).
\newblock Jail-based court notifications to improve appearance rates following
  early pretrial release.
\newblock {\em American Journal of Criminal Justice}, pages 1--21.

\bibitem[Goldkamp and White, 2006]{goldkamp2006restoring}
Goldkamp, J.~S. and White, M.~D. (2006).
\newblock Restoring accountability in pretrial release: The {P}hiladelphia
  pretrial release supervision experiments.
\newblock {\em Journal of Experimental Criminology}, 2:143--181.

\bibitem[Graef et~al., 2023]{graef2023systemic}
Graef, L., Mayson, S.~G., Ouss, A., and Stevenson, M.~T. (2023).
\newblock Systemic failure to appear in court.
\newblock {\em U. Pa. L. Rev.}, 172:1.

\bibitem[Graef and Zanger-Tishler, 2023]{graef2023consequences}
Graef, L. and Zanger-Tishler, M. (2023).
\newblock Consequences of criminal justice contact---label bias in pretrial
  risk assessments.
\newblock {\em (Working paper)}.

\bibitem[Hastings et~al., 2021]{hastings2021reducing}
Hastings, C., Thomas, C., Ostermann, M., Hyatt, J.~M., and Payne, S. (2021).
\newblock Reducing missed appointments for probation and parole supervision: a
  randomized experiment with text message reminders.
\newblock {\em Cambridge Journal of Evidence-Based Policing}, 5:170--183.

\bibitem[Holsinger et~al., 2023]{holsinger2023pretrial}
Holsinger, A.~M., Lowenkamp, C.~T., and Pratt, T.~C. (2023).
\newblock Is pretrial detention an effective deterrent? {A}n analysis of
  failure to appear and rearrest says no.
\newblock {\em Fed. Probation}, 87:3.

\bibitem[Howat et~al., 2016]{howat2016improving}
Howat, H., Forsyth, C.~J., Biggar, R., and Howat, S. (2016).
\newblock Improving court-appearance rates through court-date reminder phone
  calls.
\newblock {\em Criminal Justice Studies}, 29(1):77--87.

\bibitem[Jurva, 2021]{jurva2021impacts}
Jurva, G. (2021).
\newblock The impacts of the covid-19 pandemic on state \& local courts study
  2021: A look at remote hearings, legal technology, case backlogs, and access
  to justice.
\newblock Technical report, Thomson Reuters Institute.

\bibitem[Kofman, 2019]{kofman_2019}
Kofman, A. (2019).
\newblock Digital jail: How electronic monitoring drives defendants into debt.
\newblock {\em The New York Times Magazine}.
\newblock Accessed: 2024-02-28.
  \url{https://www.nytimes.com/2019/07/03/magazine/digital-jail-surveillance.html}.

\bibitem[Lattimore et~al., 2020]{lattimore2020prevalence}
Lattimore, P.~K., Tueller, S., Levin-Rector, A., and Witwer, A. (2020).
\newblock The prevalence of local criminal justice practices.
\newblock {\em Fed. Probation}, 84:28.

\bibitem[Loeffler and Nagin, 2022]{loeffler2022impact}
Loeffler, C.~E. and Nagin, D.~S. (2022).
\newblock The impact of incarceration on recidivism.
\newblock {\em Annual Review of Criminology}, 5:133--152.

\bibitem[Lowenkamp et~al., 2018]{lowenkamp2018assessing}
Lowenkamp, C.~T., Holsinger, A.~M., and Dierks, T. (2018).
\newblock Assessing the effects of court date notifications within pretrial
  case processing.
\newblock {\em American Journal of Criminal Justice}, 43:167--180.

\bibitem[{National Conference of State Legislatures}, 2018]{ncsl2018}
{National Conference of State Legislatures} (2018).
\newblock Pretrial release violations and bail forfeiture.
\newblock Technical report, National Conference of State Legislatures.

\bibitem[Nice, 2006]{nice2006court}
Nice, M. (2006).
\newblock Court appearance notification system: Process and outcome evaluation.
\newblock Technical report, Multnomah County Budget Office.

\bibitem[Owens and Sloan, 2023]{owens2023can}
Owens, E. and Sloan, C. (2023).
\newblock Can text messages reduce incarceration in rural and vulnerable
  populations?
\newblock {\em Journal of Policy Analysis and Management}, 42(4):992--1009.

\bibitem[Pew, 2022]{pew2022disparities}
Pew (2022).
\newblock Racial disparities persist in many {U.S.} jails.
\newblock Technical report, The Pew Charitable Trusts.

\bibitem[Pohl and Gabrielson, 2019]{sacbee}
Pohl, J. and Gabrielson, J. (2019).
\newblock We revealed {C}alifornia jails are in crisis. {G}avin {N}ewsom is
  calling for more oversight.
\newblock {\em The Sacramento Bee}.

\bibitem[Schnacke et~al., 2012]{schnacke2012increasing}
Schnacke, T.~R., Jones, M.~R., and Wilderman, D.~M. (2012).
\newblock Increasing court-appearance rates and other benefits of live-caller
  telephone court-date reminders: The {Jefferson County, Colorado}, {FTA} pilot
  project and resulting court date notification program.
\newblock {\em Court Review}, 48:86.

\bibitem[Smith, 2022]{smith2022pretrial}
Smith, S.~S. (2022).
\newblock How pretrial incarceration diminishes individuals' employment
  prospects.
\newblock {\em Fed. Probation}, 86:11.

\bibitem[Tomkins et~al., 2012]{tomkins2012experiment}
Tomkins, A.~J., Barnstein, B., Herian, M.~N., and Rosenbaum, D.~I. (2012).
\newblock An experiment in the law: Studying a technique to reduce failure to
  appear in court.
\newblock {\em Court Review}, 48:96.

\bibitem[White, 2006]{white2006court}
White, W. (2006).
\newblock Court hearing call notification project.
\newblock Technical report, Criminal Justice Coordinating Council \& Flagstaff
  Justice Court, Coconino County Arizona.

\bibitem[Zottola et~al., 2023]{zottola2023court}
Zottola, S.~A., Crozier, W.~E., Ariturk, D., and Desmarais, S.~L. (2023).
\newblock Court date reminders reduce court nonappearance: A meta-analysis.
\newblock {\em Criminology \& Public Policy}, 22(1):97--123.

\end{thebibliography}
\bibliographystyle{apalike}

\clearpage
\appendix
\renewcommand\thetable{\thesection\arabic{table}}
\renewcommand\thefigure{\thesection\arabic{figure}}
\renewcommand{\theequation}{\thesection\arabic{equation}}
\setcounter{table}{0}
\setcounter{figure}{0}
\setcounter{equation}{0}
\onehalfspacing

\section*{Appendix}

\section{Treatment Assignment}
In the first phase of the experiment
(i.e., for clients with initial court dates between \firstCourtDateFirstPhase{} and \lastCourtDateFirstPhase{}),
clients in the treatment condition received an
introductory text message up to seven days before their first court date reminder.
Occasionally, however,
court dates once eligible for reminders may have become ineligible 
in this interim period
after the introductory message was sent
(e.g.,
because the attorney indicated they would appear on the client's behalf,
or because the recipient may have opted out of text message reminders immediately after their introductory message).
As a result, 
\numClientsNoReminderInShort{} of the \numClientsShort{} clients in the treatment condition did not receive a reminder for their initially scheduled court date.
Nevertheless, we include in the treatment condition
all clients who received an introductory message, 
regardless of whether or not a reminder was actually sent, 
as the introductory text message could itself impact behavior.
In the second phase of the experiment
(i.e., for clients with initial court dates between \firstCourtDateSecondPhase{} and \lastCourtDateSecondPhase{}),
we adjusted our protocol to address this issue, 
sending the introductory message and the first court date reminder at the same time.
This change ensures that all clients in the treatment condition did in fact receive at least one reminder.

At the end of the first phase of the experiment, all clients in the first phase were transitioned to receive text messages reminders for any future court dates, regardless of whether they were initially assigned to treatment or control.
As a result, our estimate of the effect of reminders on long-term outcomes is likely conservative, since some clients in the control condition received reminders for part of the observation window.
This pattern does not affect our estimate of reminders on the issuance of bench warrants at the first court date, since that outcome is measured before any transitioning occurred.
No clients in the second phase of the experiment were transitioned,
i.e., clients in the control condition in the second phase did not receive reminders during the observation period.

\begin{figure}[t]
    \centering
    \includegraphics[width=\textwidth]{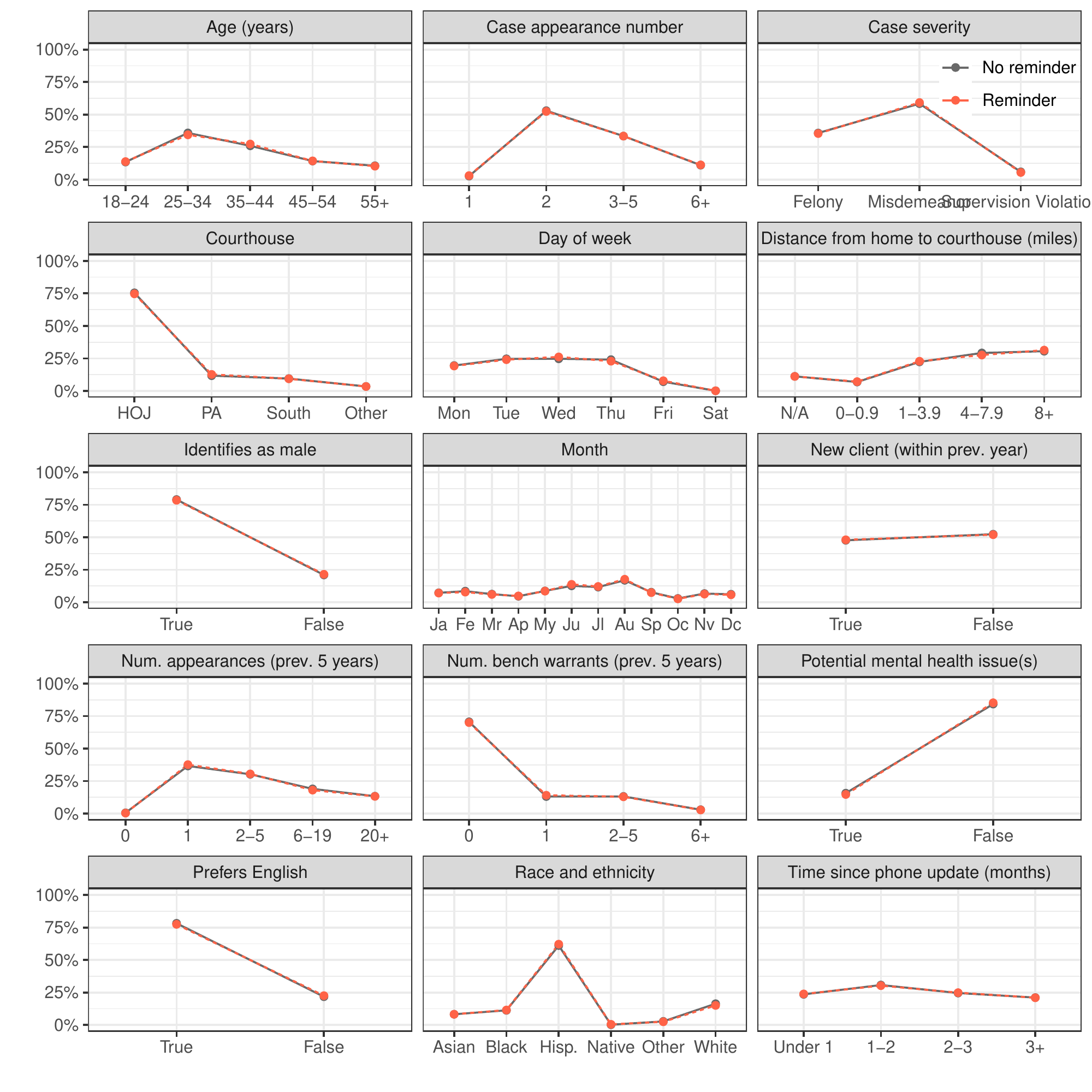}
    \caption{Covariate distributions for the treatment and control conditions were nearly identical, confirming that our assignment mechanism correctly randomly assigned clients to the two conditions.
    Statistics in this figure are drawn from the last two columns in Table~\ref{tab:pop_characteristics}.
    }
    \label{fig:full_balance_plot}
\end{figure}

To confirm that our assignment procedure indeed randomly assigned clients to treatment or control, we examined balance plots (Table~\ref{tab:pop_characteristics} and Figure~\ref{fig:full_balance_plot}).
Across a wide range of covariates, we see that the distributions are nearly identical between the two conditions, as expected.

\clearpage
\section{Client Population Distributions}

\begin{longtable}{lLLLLL}
\sampleStatsTable
\caption{
Population characteristics for five subsets of SCCPDO clients. 
The first population, ``all'' clients, was created by considering all SCCPDO clients with a reminder-eligible court date during the experiment window and measuring client and case characteristics at the first reminder-eligible court date for each client within this window.
The second population, ``cell on file'', was constructed by considering all SCCPDO clients who had a cellphone number on file and a reminder-eligible court date during the experiment window,
and measuring attributes at the first reminder-eligible court date for each client within the experiment window.
The third population, ``experiment'' clients, represents all clients in the experiment population at their first observed court date.
The fourth and fifth populations, ``treatment'' clients and ``control'' clients, 
further break down the experiment population by their random assignment.
Statistics for the last two populations are visually represented in Figures~\ref{fig:limited_balance_plot} and \ref{fig:full_balance_plot}.
}
\label{tab:pop_characteristics}
\end{longtable}

\clearpage
\section{Logistic Regression Coefficient Estimates}

\begin{longtable}{llll}
\coefficientEstimatesTable
\caption{
Logistic regression coefficient estimates for the three covariate-adjusted models described in Section~\ref{sec:results}. 
Model numbers correspond to those listed in Table~\ref{tab:regression_results}. 
Coefficient estimates are on the log-odds scale, with standard errors in parentheses. 
A single star indicates that the corresponding logistic regression coefficient estimate has a p-value
between 0.01 and 0.05,
a double star indicates a p-value between 0.001 and 0.01,
and a triple star indicates p-values under 0.001.
}
\label{tab:model_coefs}
\end{longtable}

\clearpage
\section{Time from Bench Warrant to Incarceration}

\begin{figure}[h]
    \centering
    \includegraphics[width=0.7\columnwidth]{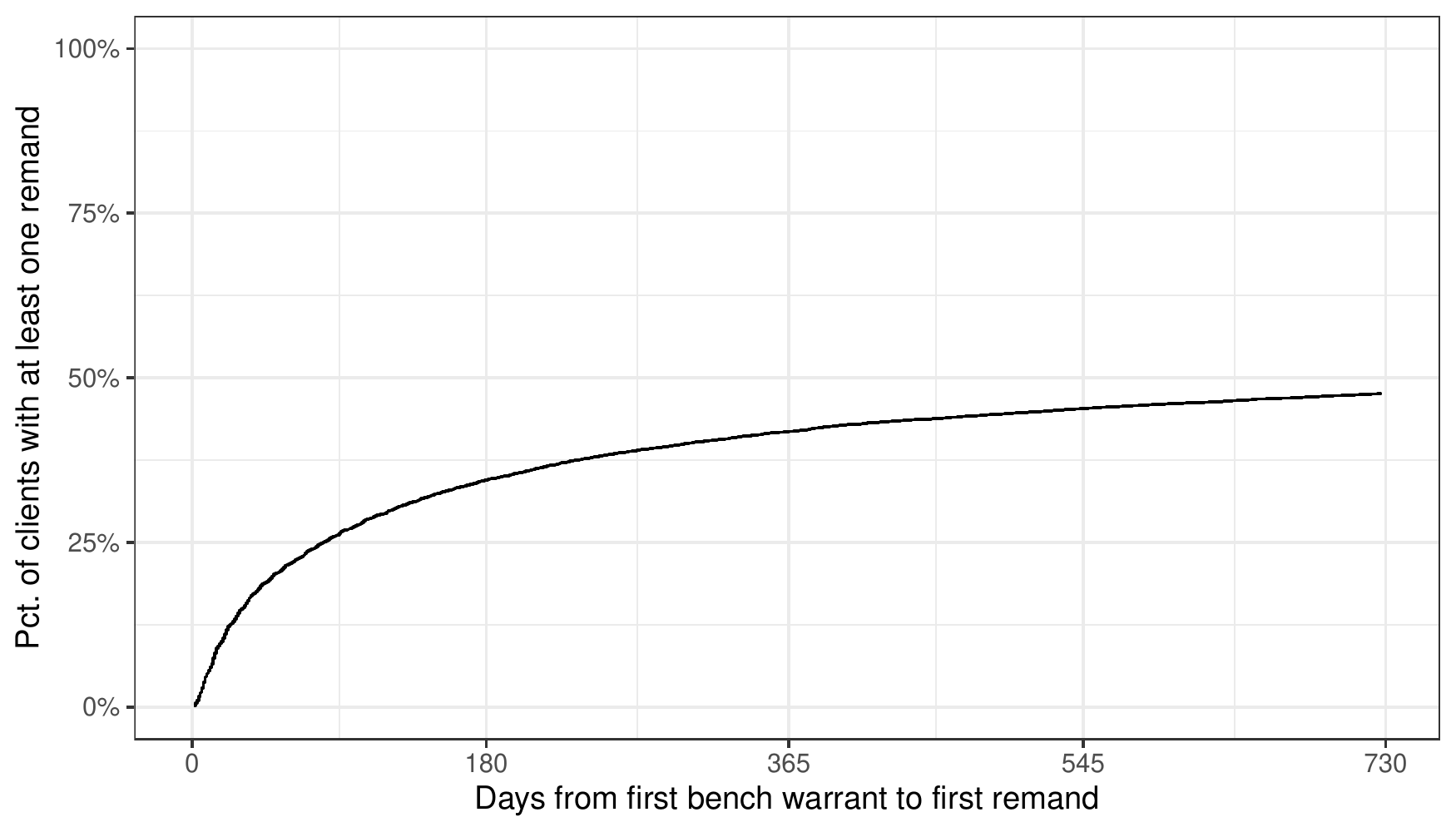}
    \caption{Empirical cumulative distribution of the number of days it takes for SCCPDO clients to be remanded to jail after receiving a bench warrant.
    This distribution was calculated for all SCCPDO clients who received at least one bench warrant between January 2019 and October 2021.
    Days to incarceration are measured relative to clients' first bench warrant received during this period.
    Roughly half of SCCPDO clients were remanded to custody within two years of their first bench warrant.}
    \label{fig:time-to-remand}
\end{figure}

\section{Confirmation and Appearance Behavior}

\begin{table}[h]
{\centering
\begin{tabularx}{0.6\linewidth}{lYY}
\toprule
Confirmed? & Proportion of Clients & Bench Warrant Rate \\
\midrule
Yes & \pctShortConfirmed{} & \bwRateShortConfirmed{} \\
No & \pctShortUnconfirmed{} & \bwRateShortUnconfirmed{} \\
\bottomrule
\end{tabularx}
\caption{Proportion of clients in treatment arm who confirmed and did not confirm, and their corresponding bench warrant rates at the first observed court date. Note that confirmation behavior is a response to the reminder, and as such the act of confirmation cannot be interpreted causally, since this interpretation would be susceptible to post-treatment bias.}
\label{tab:confirmation_and_appearance_behavior}}
\end{table}

\clearpage
\section{Spanish and Vietnamese Reminder Examples}

\begin{figure}[h]
    \centering
    \includegraphics[width=0.92\columnwidth]{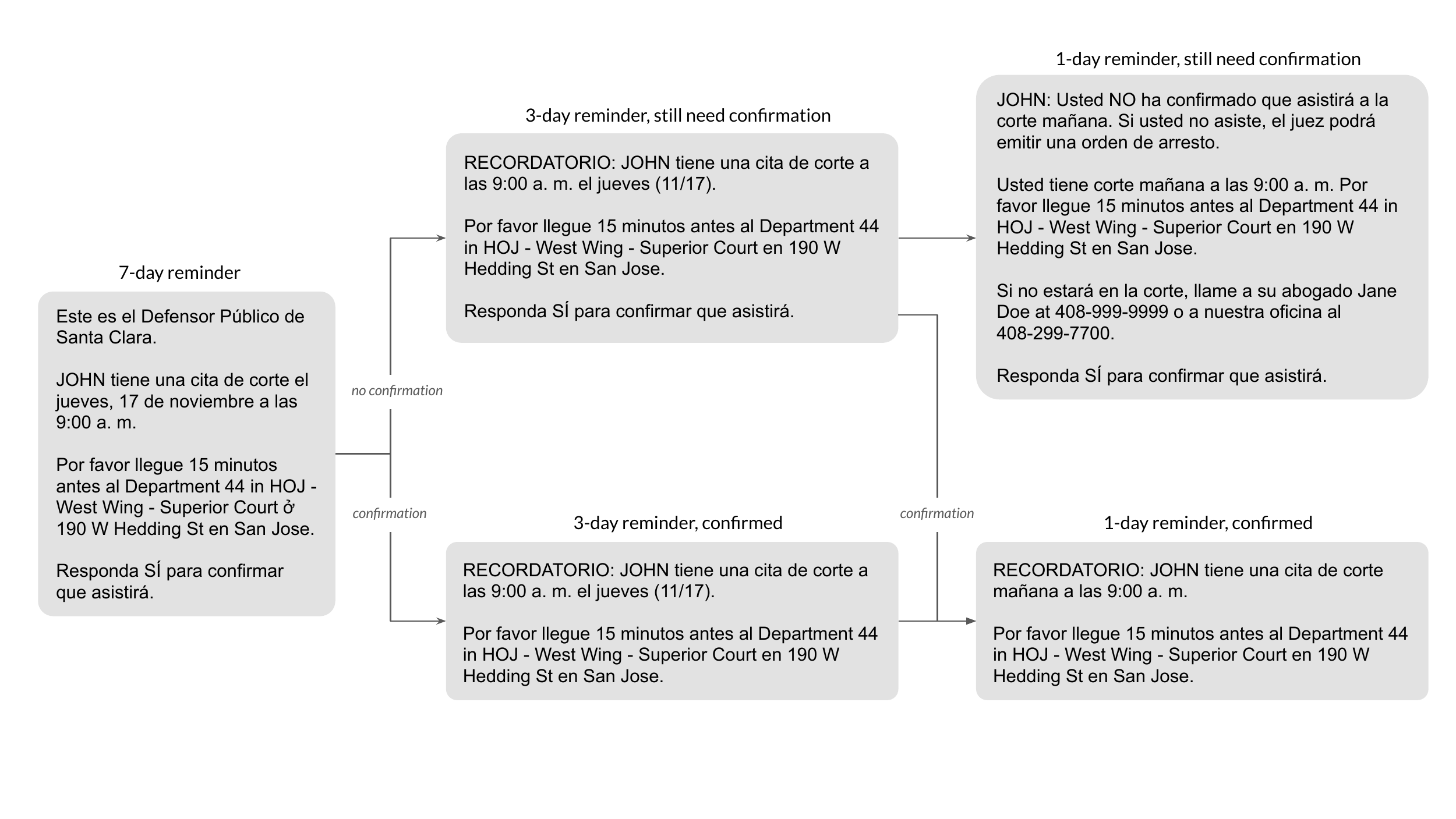}
    \vspace{-2.5em}
    \caption{Example of reminder flow in Spanish.}
    \label{fig:treatment-diagram-spanish}
\end{figure}

\begin{figure}[h]
    \centering
    \includegraphics[width=0.92\columnwidth]{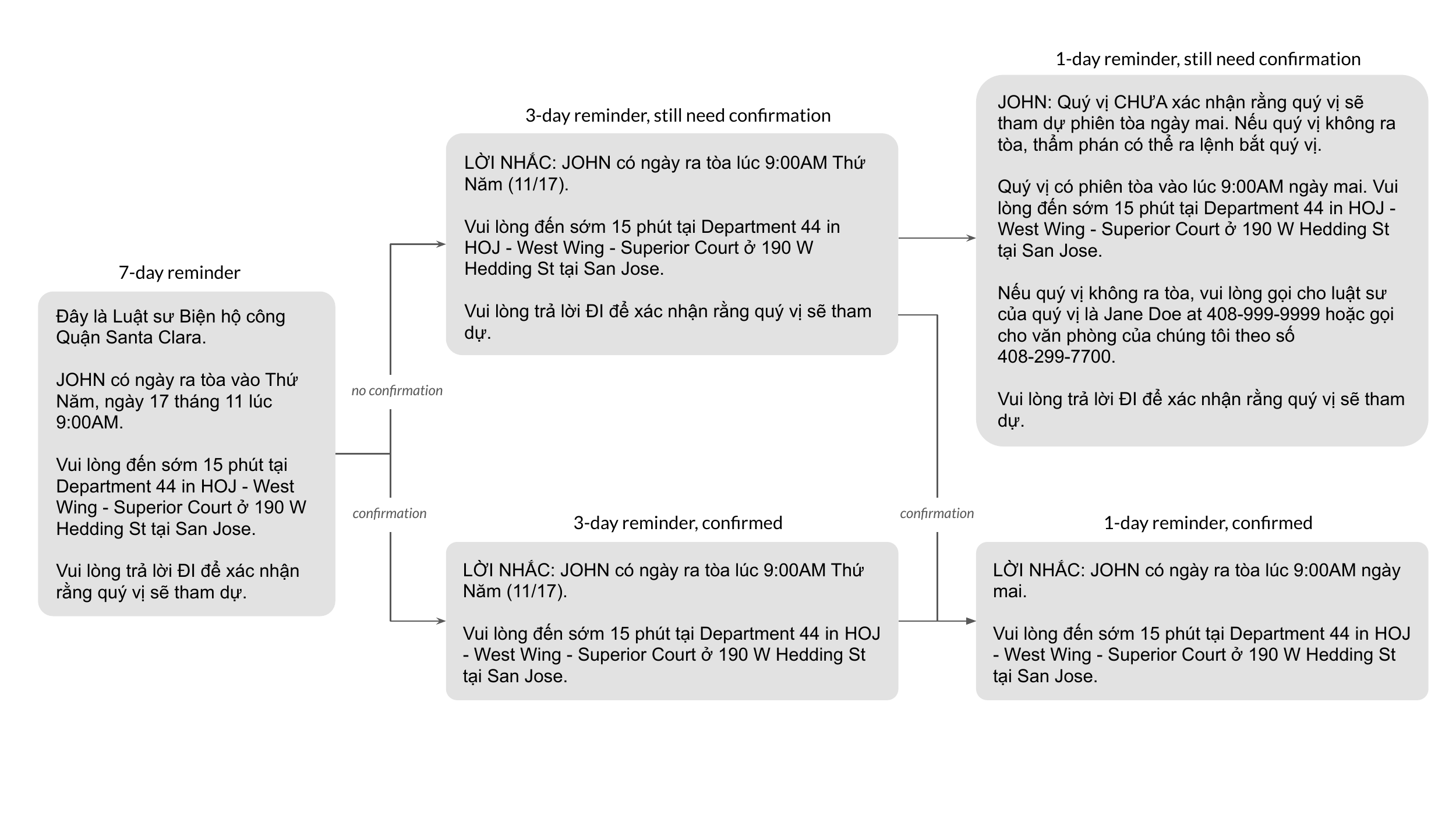}
    \vspace{-2.5em}
    \caption{Example of reminder flow in Vietnamese.}
    \label{fig:treatment-diagram-vietnamese}
\end{figure}

\end{document}